\documentstyle[11pt]{article}
\begin{document}
\newcommand{\ECM}{\em Departament d'Estructura i Constituents de la
Mat\`eria
                  \\ Facultat de F\'\i sica, Universitat de Barcelona \\
                     Diagonal 647, 08028 Barcelona, Spain}
\newcommand{\NPB}[3]{{\em Nucl.Phys.} {\bf B#1}(19{#2}){#3}}
\newcommand{\PLB}[3]{{\em Phys.Lett.} {\bf B#1}(19{#2}){#3}}
\newcommand{\IJA}[3]{{\em Int.J.Mod.Phys.} {\bf A#1}(19{#2}){#3}}
\def\QQa{\renewcommand{\baselinestretch}{1.3}\Huge\large\normalsize}
\def\a{\alpha}  \def\b{\beta} \def\g{\gamma} \def\G{\Gamma}
\def\d{\delta} \def\D{\Delta} \def\e{\epsilon} \def\ee{\varepsilon}
\def\z{\zeta} \def\th{\theta} \def\TH{\Theta} \def\tth{\vartheta}
\def\k{\kappa} \def\l{\lambda} \def\L{\Lambda} \def\m{\mu} \def\n{\nu}
\def\cs{\xi} \def\Cs{\Xi} \def\p{\pi} \def\P{\Pi} \def\r{\rho} \def\s{\sigma}
\def\S{\Sigma} \def\t{\tau} \def\y{\upsilon} \def\Y{\upsilon}
\def\f{\phi} \def\F{\Phi} \def\x{\chi} \def\ps{\psi} \def\Ps{\Psi}
\def\o{\omega} \def\O{\Omega} \def\vf{\varphi}
\def\pa{\partial} \def\da{\dagger} \def\dda{\ddagger}

\def\ph4{\lambda\varphi^{4}} \def\bfnc{$\beta$-function}
\def\dr{differential renormalization}
\def\vs{\vspace{.25in}}
\def\thefootnote{\fnsymbol{footnote}}
\pagestyle{empty}
{\hfill \parbox{6cm}{\begin{center} UB-ECM-PF 92/5\\
                                    March 1992
                     \end{center}}}
\vspace{1.5cm}

\begin{center}
\large{\bf DIFFERENTIAL RENORMALIZATION OF \\MASSIVE QUANTUM
FIELD THEORIES}
\vskip .6truein
\centerline {Peter E. Haagensen\footnote{e-mail: HAGENSEN@EBUBECM1}
and Jos\'e I. Latorre\footnote{e-mail: LATORRE@EBUBECM1}}
\end{center}
\vspace{.3cm}
\begin{center}
\ECM
\end{center}
\vspace{1.5cm}

\centerline{\bf Abstract}
\medskip

We extend the method of \dr\ to massive quantum field theories,
treating in particular $\ph4$-theory and QED.
As in the massless case, the method proves to be simple and powerful,
and we are able to find, in particular, compact explicit coordinate
space expressions
for the finite parts of two notably complicated diagrams,
namely, the 2-loop 2-point function in $\ph4$ and the 1-loop vertex
in QED.

\newpage
\pagestyle{plain}
\QQa

Differential renormalization (DR)[1] is a coordinate space
renormalization procedure which removes the divergences of bare
amplitudes by writing these as derivatives of less singular functions
and then prescribing the derivatives to be integrated by parts with
all surface terms discarded. The singularities of amplitudes that did
not allow for Fourier transformation into momentum space are eliminated
and one ends up with finite, renormalized amplitudes that satisfy
renormalization group equations. The standard and simplest example of
the procedure is for the 1-loop 4-point ``bubble" diagram in massless
$\ph4$, where the following identity is used:

\begin{equation}\label{bubble}
{1\over x^4} = -{1\over 4} \Box{\ln x^2M^2\over x^2} \qquad x\not=
0.\end{equation}

\noindent $M$ is an integration constant which will become the
subtraction scale of the renormalized amplitude. If we prescribe the
laplacian to act only after integration by parts, then the r.h.s. above
has a finite Fourier transform, and in fact {\em defines} the
renormalized value of the (divergent) Fourier transform of the l.h.s..
Generally, one can say that differential renormalization provides a
prescription to continue distributions defined almost everywhere (i.e.,
bare amplitudes, undefined at a finite number of singular points) into
{\sl bona fide} distributions defined everywhere (renormalized
amplitudes).

The procedure exemplified above has been carried out at higher loops and
in different models [1,2,3]. One can furthermore check explicitly the
form of the divergences being subtracted by examining the surface terms
which contain them, thus verifying that the method indeed corresponds to
a counterterm subtraction procedure [4].

In this Letter, we extend the method to allow for the inclusion of
masses in an exact treatment. At the outset, it is clear that the
presence of masses should not interfere with the method since the
renormalization procedure is related to short-distance singularities
whereas masses only change the long-distance behavior of correlators.
In fact, one could simply take a pragmatic attitude and expand all
massive propagators around zero mass and then proceed to use standard
massless \dr . However, the question remains whether massive field
theories are amenable to a  treatment {\em exact} in the mass parameter
at each order in perturbation theory. We find
that our expectations are entirely fulfilled. In analogy to the massless
case, renormalization is accomplished through the use of (massive)
\dr\ identities and an integration by parts prescription. We treat QED
at one loop and $\ph4$-theory at two loops and, most notably, we are
able to give in closed form the full renormalized expressions for the
2-loop ``setting sun" diagram of $\ph4$ and the 1-loop vertex of QED,
two results of considerable difficulty of calculation in standard
momentum space treatments. In the end, we shall discuss the general
features of our {\em modus operandi}.

First of all, let us recall the euclidean  propagator for a massive
scalar particle in four dimensions,
\begin{equation}
\Delta(x,m) = {1\over 4\pi^2} {mK_1(mx)\over x}
\end{equation}
where $K_1$ is a modified Bessel function. Up to permutations of external
legs the only 1-loop diagram contributing to the 4-point function in
$\ph4$ theory is:
\begin{equation}
\Gamma^{(4)}_{bare}(x) = {\lambda^2\over 2} \left({1\over 4\pi^2}
{mK_1(mx)\over x} \right)^2  \end{equation}
At short distances, this has the same log divergence as the
corresponding massless diagram; in the spirit of differential
renormalization we look for an expression that corresponds to the
massive generalization of Eq.(1). We find:

\begin{equation}
{m^2K_1^2(mx)\over x^2}={1\over 2}(\Box -4m^2){mK_0(mx)K_1(mx)\over x}+
\pi^2\ln{{\bar M}^2\over m^2}\d^{(4)}(x)
\end{equation}
where ${\bar M}\equiv 2 M/\gamma$ and $\gamma= 1.781072...$ is the Euler
constant. The contact term has been added in order to give a
well-defined massless limit to the r.h.s. above, coinciding in fact with
Eq.(1). We shall further analyze this later on, and in fact we will see
that Eq.(4) can be seen as a prototype in our treatment, since it
presents all the basic guiding elements in working out massive DR
identities in general. The renormalized 1-loop 4-point function,
$\G_R^{(4)}(x,M)$, is gotten by simply substituting the above in Eq.(3).

The one-loop beta function can be obtained from the renormalization
group equation
\begin{equation}
\left( M{\pa\over\pa M}+\b {\pa\over\pa\l}+\g_mm^2{\pa\over\pa m^2}-
4\g\right) [-\l\d^{(4)}(x)+\G^{(4)}_R(x,M)]=0,
\end{equation}
where $\b (\l )$ is the \bfnc , $\g_m(\l )$ is the anomalous mass
dimension, and $\g (\l )$ the anomalous dimension of $\vf$.
Dropping terms of higher order in $\l$ (viz., $\g_m$ and $\g$), the
result for the \bfnc\ is:
\begin{equation}
\b (\l )=3\left({\l\over 16\p}\right)^2
\end{equation}
with the three permutations of the $s$, $t$ and $u$ channels having
been added. Our result for the bubble diagram can be easily Fourier
transformed. The result in momentum space is:
\begin{equation}
{\tilde\Gamma}_{R}^{(4)}(p) = {\l^2\over 32\p^2}\left[
\ln{{\bar M}^2\over m^2}-\sqrt{1+{4m^2\over p^2}}\ln{\sqrt{1+{4m^2\over
p^2}}+1\over\sqrt{1+{4m^2\over p^2}}-1}\right]
\end{equation}
which agrees with the standard result in textbooks [5]. Note the
appearance of the 2-particle threshold $4m^2$ in the operator which was
extracted in Eq.(4). This will be a recurring feature in what follows.
We have also worked out this amplitude for two different
masses running along the two lines in the loop; we do not present it
here, but rather just mention that the result generalizes
the one above and, as expected, the threshold $(m_1+m_2)^2$ appears
instead of $4m^2$.

A far less trivial example showing the power of \dr~ is given by the
computation of the 2-loop correction to the 2-point function (often
called the ``setting sun"). The complexity of the finite parts of this
diagram is such that they are not presented in standard reviews of
$\ph4$ renormalization. The bare expression for the diagram is:
\begin{equation}
\Gamma^{(2)}_{bare}(x)= {\lambda^2\over 6}
\left( {1\over 4\pi^2}{mK_1(mx)\over x}\right)^3.
\end{equation}
This bare amplitude is quadratically divergent and thus requires
the extraction of two laplacians in order to have a good Fourier
transform. Little effort is needed to verify the identity that leads to
the following renormalized value of the above diagram:
\begin{eqnarray}
\G^{(2)}_R(x,M)&=&{\l^2\over 96(4\p^2)^3}
\left[(\Box -9m^2)(\Box -m^2)\left(m^2K_0(mx)K_1^2(mx)+m^2
K_0^3(mx) \right)\right. \nonumber\\
&&\left. +2\p^2\ln{\bar{M}^2\over m^2}(\Box +am^2)\d^{(4)}(x)\right]
\end{eqnarray}
This result is by itself a remarkable application of \dr . It
corresponds to a closed and compact expression for the
renormalized 2-loop 2-point function. Its Fourier transform is
complicated and we will not work it out here. Again, we note the
appearance of the 3-particle production threshold as a healthy sign of
the procedure. Further, the second operator takes the
form $(\Box -m^2)$, which vanishes on mass shell. The coefficient of
the $\Box \delta^4(x)$ term is fixed as in the previous example by the
requirement of a smooth massless limit and this, in turn, fixes
$\gamma(\lambda)$ in the renormalization group equation to its standard
value
\begin{equation}
\g (\l )={1\over 12}\left({\l\over 16\p}\right)^2.
\end{equation}
On the other hand, the coefficient $a$ of $m^2\d^{(4)}(x)$ is not fixed by
a strict massless limit, which is to be expected, since it corresponds
to a mass subtraction, and fixes the (scheme-dependent) $\g_m$ function
in the renormalization group equation. In particular, $a$ can be
chosen to be $-1$ so that the whole 2-loop contribution vanishes
on mass shell.

Though we do not intend to present an exhaustive computation of the
perturbative expansion of massive $\ph4$, let us comment on a few more
diagrams. There are only two more diagrams contributing at two loops to
the 4-point function. The first one, usually called the ``ice cream
cone", though extremely difficult in momentum space, is easily
calculated along the lines we have sketched above. The second one,
consisting of two 4-point 1-loop bubbles attached together (and thus,
the ``double bubble") is not as straightforward due to the presence of a
convolution.

We now turn to the computation of 1-loop diagrams in QED
\footnote{For a complete set of momentum space techniques to compute
one-loop diagrams in gauge theories, see [6].}.
The (massive) fermion and photon propagators are, respectively:
\begin{equation} S(x,m)={1\over 4\p^2}(\pa\!\!\! /-m){mK_1(mx)\over x},
\end{equation}
and
\begin{equation}
\D_{\m\n}(x)={1\over 4\p^2}{\d_{\m\n}\over x^2},
\end{equation}
this latter one being given in Feynman gauge. The $\bar{\ps}A\!\!\!/\ps$
vertex has the value $ie\g_\m$. The 1-loop fermion self-energy then reads:
\begin{equation}
\S_{bare}(x)=\left({ie\over 4\p }\right)^2{1\over x^2}\g_\m
(\pa\!\!\! /-m){mK_1(mx)\over x} \g_\m .
\end{equation}
The mass piece above has a logarithmic short-distance divergence, while
the derivative term is linearly divergent at short distances. These
divergences are eliminated by the following massive DR identity:
\begin{equation}
{mK_1(mx)\over x^3}={1\over 2}(\Box -m^2){K_0(mx)\over x^2}+
\p^2\ln{\bar{M}^2 \over m^2}\d^{(4)}(x),
\end{equation}
where again the contact term is determined such as to give a
well-defined massless limit to the r.h.s. (and again equal to the
corresponding massless identity). The final, renormalized expression for
the fermion self-energy will then be:
\begin{eqnarray}
\S_R (x,M)&=&{e^2\over 32\p^2}\left[ (\Box -m^2)\left( (\pa\!\!\!/+4m)
{K_0(mx)\over x^2}+{m^2\over 2}{x\!\!\!/ K_0(mx)\over
x^2}\right)\right. \nonumber\\
&&\left. +2\p^2\ln{\bar{M}^2 \over m^2}(\pa\!\!\!/+bm)\d^{(4)}(x)\right].
\end{eqnarray}
The non-local piece of this amplitude vanishes identically on
mass-shell due to the operator $(\Box -m^2)$, as expected. In
analogy
to the previous case, the $\pa\!\!\!/\delta^4(x)$ contact term is fixed
by the massless limit (and leads to the standard value
for the anomalous dimension of the fermionic field), whereas
the $m\delta^4(x)$ term is not. We leave the coefficient $b$
undetermined, and again it is clear that a particular choice, viz.
$b=-1$, makes the whole amplitude vanish on mass-shell.
The Fourier-transformed, momentum space amplitude can be obtained fairly
straightforwardly, and we do not present it here.

The bare 1-loop vacuum polarization is:
\begin{equation}
\P_{\m\n}^{bare}(x)=-\left({ie\over 4\p^2}\right)^2 tr\left(
\g_\m (\pa\!\!\!/-m){mK_1(mx)\over x}\g_\n (-\pa\!\!\!/-m){mK_1(mx)\over
x} \right).
\end{equation}
By standard manipulations with Bessel functions and the use of a
massive DR identity, Eq. (4), we find the following renormalized vacuum
polarization at one loop:
\begin{eqnarray}
\P_{R\m\n}(x,M)&=&-{e^2\over 24\p^4}(\pa_\m\pa_\n -\d_{\m\n}\Box)\left[
(\Box -4m^2)\left({mK_0(mx)K_1(mx)\over x}
\right.\right. \nonumber\\
&&\left.\left. +{m^2\over 2}(K_1^2(mx)- K_0^2(mx))\right)
+2\p^2\ln{\bar{M}^2 \over m^2}\d^{(4)}(x)\right].
\end{eqnarray}
As expected, the result is automatically transverse and shows the
presence of the 2-particle production threshold in the
operator coming in front of the non-local piece of the amplitude.

We finally turn to the 1-loop vertex. Its bare value is:
\begin{equation}
V_\m^{bare}(x,y)=\left({ie\over 4\p^2}\right)^3
\g_\r\, (\pa\!\!\! /-m){mK_1(mx)\over x}\,\g_\m\, (-\pa\!\!\! /-m)
{mK_1(my)\over y}\,\g_\r\, {1\over (x-y)^2}.
\end{equation}
In the above, the only divergent piece is the one containing the
two derivatives:
\begin{equation}
V_\m^{div}(x,y)=-\left({ie\over 4\p^2}\right)^3
\g_\r\, \pa\!\!\! / {mK_1(mx)\over x}\,\g_\m\, \pa\!\!\! /
{mK_1(my)\over y}\,\g_\r\, {1\over (x-y)^2}.
\end{equation}
We renormalize this by an identical procedure used in the massless case
[1], i.e., by integrating the two derivatives by parts onto $1/(x-y)^2$,
and separating that into a (finite) traceless piece and a (divergent)
trace piece. The divergence, thus isolated in the trace, is then
renormalized with massive DR identity Eq.(4). The final result reads:
\begin{eqnarray}
&&V_\m^R(x,y,M)=\left({ie\over 4\p^2}\right)^3\left[ 2\g_b\g_\m\g_a
\left({\pa\over\pa x^a}\left[{mK_1(mx)\over x}{\pa\over\pa y^b}\left(
{mK_1(my)\over y}\right){1\over (x-y)^2}\right]\right.\right.
\nonumber\\
&&\left. -{\pa\over\pa y^b}\left[{m^2K_1(mx)K_1(my)\over x y}
{\pa\over\pa x^a}{1\over (x-y)^2}\right]
-{m^2K_1(mx)K_1(my)\over x y}(\pa_a\pa_b -
{1\over 4}\d_{ab}\Box){1\over (x-y)^2}\right)\nonumber\\
&&-2\p^2\g_\m (\Box -4m^2)
{mK_0(mx)K_1(mx)\over x}\d^{(4)}(x-y)
-4\p^4\g_\m\ln {\bar{M}^2\over m^2}\d^{(4)}(x)\d^{(4)}(x-y)\nonumber\\
&&\left. +4m{1\over (x-y)^2}\left({\pa\over\pa y^\m}-{\pa\over\pa x^\m}-
{m\over 2}\g_\m\right)\left({m^2K_1(mx)K_1(my)\over xy}\right)
\right].
\end{eqnarray}

This concludes our results. We have extended the method of \dr\ to
massive $\ph4$ and QED, and the results we present here, though not
intended to be exhaustive, should indicate both the feasibility of, and
the guidelines for, the treatment of massive fields at higher loops and
in different models. These guidelines basically are: a) the presence of
an equal number of K-functions on both sides of massive DR identities;
b) a direct connection to massless \dr , through the substitution of
$\ln x^2M^2$ in this latter case for $K_0(mx)$; c) the appearance of the
appropriate particle production thresholds in the differential operators
present in massive DR identities, and d) the appearance of contact
terms, due to the requirement of existence of the massless limit for
renormalized distributions. These features have been of central
importance in helping us find all the renormalized amplitudes presented
here.

Two final comments are in order regarding contact terms in massive DR
identities. Firstly, the presence of these terms may seem puzzling,
since they vanish at points where massive DR identities are valid
mathematical equations, i.e., for $x\not= 0$. However, insofar as these
identities are prescriptions for extending (i.e., renormalizing)
distributions, they should be understood as identities  everywhere,
including at contact, and specific contact terms will then determine
precisely and uniquely the value of the extended (i.e., renormalized)
distributions at those points. Finally, it is also worth noting that
contact terms encode the entire renormalization group freedom present in
the renormalization procedure. Our choice has been to fix contact terms
in order to agree with the massless case, which corresponds to a
wave function renormalization prescription, and to leave undetermined
the mass subtraction, though an on-mass-shell scheme has been indicated.
\vs

{\em Acknowledgments -} We would like to thank D.Z. Freedman for
ongoing discussions. This work was supported in part by CAICYT grant
AEN90-0033 and by the EEC Science Twinning Grant SCI-000337. P.H.
also acknowledges a grant from the Ministerio de Educaci\'on y
Ciencia, Spain.\vs

{\em References}\\

\noindent
{\bf [1]} Freedman, D.Z., K. Johnson and J.I. Latorre, {\em Nucl.Phys.}
{\bf B}, in press.\\
{\bf [2]} Haagensen, P.E., {\em Mod.Phys.Lett.} {\bf A7}(1992)893.\\
{\bf [3]} Freedman, D.Z., G. Grignani, K. Johnson and N. Rius,
``Conformal Symmetry and\\
\mbox{}\phantom{[1]x}Differential Regularization of the 3-Gluon
Vertex'', MIT preprint CTP\#1991.\\
{\bf [4]} Freedman, D.Z., R. Mu\~noz-Tapia and X.
Vilas{\'{\i}}s-Cardona, manuscript in\\
\mbox{}\phantom{[1]x}preparation.\\
{\bf [5]} Ramond, P., {\em Field Theory, A Modern Primer}, 2nd ed.,
Addison-Wesley(1990).\\
{\bf [6]} 't Hooft, G. and M. Veltman, {\em Nucl.Phys.}
{\bf B}153(1979)365.\\
 \end{document}